\documentstyle[twocolumn,aps,floats,amssymb,epsfig]{revtex}
\begin{document}

\def\d{{\rm d}}
\def\e{{\rm e}}
\def\i{{\rm i}}
\def\O{{\rm O}}
\def\half{\mbox{$\frac12$}}
\def\eref#1{(\protect\ref{#1})}
\def\etal{{\it{}et al.}}
\def\Li{\mathop{\rm Li}}
\def\av#1{\left\langle#1\right\rangle}
\def\set#1{\left\lbrace#1\right\rbrace}
\def\stirling#1#2{\Bigl\lbrace{#1\atop#2}\Bigr\rbrace}

\draft
\tolerance = 10000

\renewcommand{\topfraction}{0.9}
\renewcommand{\textfraction}{0.1}
\renewcommand{\floatpagefraction}{0.9}
\setlength{\tabcolsep}{4pt}

\twocolumn[\hsize\textwidth\columnwidth\hsize\csname @twocolumnfalse\endcsname

\title{Power-Law Scaling for Communication Networks with Transmission Errors}
\author{ Jiann-Shing Lih and Jyh-Long Chern}
\address{ Department of Physics, National Cheng Kung University, Tainan, Taiwan 70101,
Republic of China}
\date{manuscript, 29 August, 2001}
\maketitle

\begin{abstract}
A generic communication model of a boolean network with
transmission errors is proposed to explore the power-law scaling
of states' evolution in small-world networks. In the model, the
power spectrum of the population difference between agents with
"1" and "-1" exhibits a power-law tail: $P(f)\sim f^{-\alpha}$.
The exponent $\alpha$ does not depend explicitly on the error rate
but on the structure of the networks. Error rate enters only into
the frequency scales and thus governs the frequency range over
which these power laws can be observed. On the other hand, the
exponent $\alpha$ reveals the intricate internal structure of
networks and can serve as a structure factor to classify complex
networks. The finite transmission error is shown to provide a
mechanism for the common occurrence of the power-law relation in
complex systems, such as financial markets and opinion formation.
\newline
\end{abstract}

\vskip0.1in ]

\newpage

\noindent The dynamics of complex networks are associated with
their structures and the interactions among their elements. In
complex systems, elements are often organized by communicating
information to one another with their specific links. For example,
society is organized as a social web, whose nodes are individuals
and links represent various social interactions~\cite{rf1}, and
the WWW forms a complex web whose nodes are documents and
links are URLs~\cite{rf2}. In our daily life, we communicate with
each other via language, whose nodes are words, and the links
between words are established by grammar~\cite{rf3}. The webs in
which the constituents interact bring out the complexity of those
systems. Pioneering work done by Watts and Strogatz~\cite{rf4} has
demonstrated that many natural, social, and technological networks
have topological properties in between those of regular lattices
and random graphs. These complex networks exhibit a certain degree
of clustering like regular lattices and a logarithmic increase in
the average shortest path between two nodes with the number of
nodes, as in random graphs. Watts and Strogatz proposed the
small-world model~\cite{rf4} - an interpolation between regular
lattices and random graphs - to model these network. The two
features of small-world networks lead to cluster formation and the
spread of global information through these complex systems.
\bigskip

\noindent Besides the network structure, another important factor
attributed to the behavior of complex systems is how correctly
information can be transmitted between nodes. For example, in a
social structure, the individual's attitude toward some issue is
often influenced by his/her friends and public opinion~\cite{rf5}.
Most people are inclined to the opinion of the majority, which
inclination is also called \textit{herd
behavior}~\cite{rf6,rf7,rf8}. In a perfect communication network
without information error, the states of nodes will interact and
finally evolve to a steady state. However, information transmitted
between nodes is often liable to error due to inevitable noises.
Thus, the imperfection of communication in networks always
perturbs the whole system into non-equilibrium state. Consequently,
it is natural to ask how transmission errors impact the dynamical
behavior of networks. Communication errors can be generally
classified into three groups~\cite{rf9}: (a)\textit{Technical
problem}: how accurately can the symbols of communication be
transmitted? (b)\textit{Semantic problem}: how precisely do the
transmitted symbols convey the desired meaning? and (c)
\textit{Effectiveness problem}: how effectively does the received
meaning affect conduct in the desired way? This work focuses
on the effect of the technical problem on communication
networks; namely, how accurately information can be transmitted
and what is its influence? \bigskip

\noindent  In this study, a system of \textit{N} agents is considered.
The agents are represented by nodes in a small-world structure. For simplicity,
the network model adopted here is a boolean network in which the state
of agent \textit{i} is represented by $S_{i}(t)$ = \{-1,1\}. The
evolution of the $i_{th}$ agent's state, $S_{i}(t)$, follows the updating
rule,
\begin{equation}
   S_{i}(t+1)=sgn\{\sum^{N}_{j=1}w_{ij}P_{j\rightarrow i}[S_{j}(t),\phi]\},
\end{equation}where $ w_{ij}=1(0)$ if agents \textit{i} and \textit{j}
are (not) connected. The probability function, $P_{j\rightarrow i
}[S_{j}(t),\phi]$, determines how accurately information can be
transmitted from agent \textit{j} to \textit{i}. Thus, the
transmitted information from agent \textit{j} to \textit{i} is
$S_{j}(t)$ with probability $(1-\phi)$ where $\phi$ is the
transmission error rate. Function sgn(x)=+1 (-1) if the argument
$x>0$ ($<0$). If x=0, $S_{i}(t+1)=S_{i}(t)$, that is, the transmitted
information remains unchanged. The updating rule followed here
corresponds to the majority rule~\cite{rf5}, that individual
opinion often follow public opinion. For simplicity,
this study does not consider the free will of the individual, thus
the effect of transmission errors on the dynamics of
information spreading in networks can be fully retrieved. The
evolution of the system is characterized by a succession of
discrete events $S(1), S(2), S(3),...$, where
$S(t)=\sum^{N}_{j=1}S_{j}(t)$ corresponds to the population
difference between the states +1 and -1. When $S(t) = N(-N)$, all
agents are in the same state, +1(-1), whereas $S(t)$ = 0
corresponds to an equal population of states +1 and -1. As time
passes, clusters consisting of agents with the same state begin to
develop. Accordingly, in this model the interaction between the agents
can be divided into two classes: (a) \textit{boundary} interaction
between clusters of different states and (b) \textit{internal}
interaction among the agents of the same cluster. These
two interactions compete against each other in a new way, and the
competition between internal and boundary interactions is numerically
shown to result in a nontrivial dynamical behavior - \textit{power-law
dependence on the frequency}.\bigskip

\noindent A standard Monte Carlo simulation is performed with simultaneous
updating to investigate the model presented here. The simulation follows
the procedure described by Newman and Watts~\cite{rf10}, to
construct the small-world network. Initially, each node is
connected to all of its neighbors up to some fixed $k$ range to
make a network with coordination number $z=2k$. Next, $m$
shortcuts are added between randomly selected pairs of nodes. The
randomness, $p$, of the small-world network is defined as
$p=\frac{m}{Nk}$ where $N$ is the population of the agents in the
small-world network. Although the definition of randomness adopted
here varies from the definition of randomness,$R$, in Ref.~\cite{rf4},
these two definitions ($p$ and $R$) are approximately equivalent in the
small-world regime. The simulation mainly used $N$=500, but a simulation
with $N$=5000 was performed for confirmation. The simulation began in an
initial "all +1" state, in which every nodes is set initially to the
same state, +1. In particular, the coarsening process~\cite{rf15} from the
homogeneous initial condition is heuristically investigated. After the
transient, the system evolves around some level of population difference.
However,  the observed power-law scaling relation was still attained from
the wholly random initial condition.\bigskip

\begin{figure}
\begin{center}
\includegraphics*[scale=0.6]{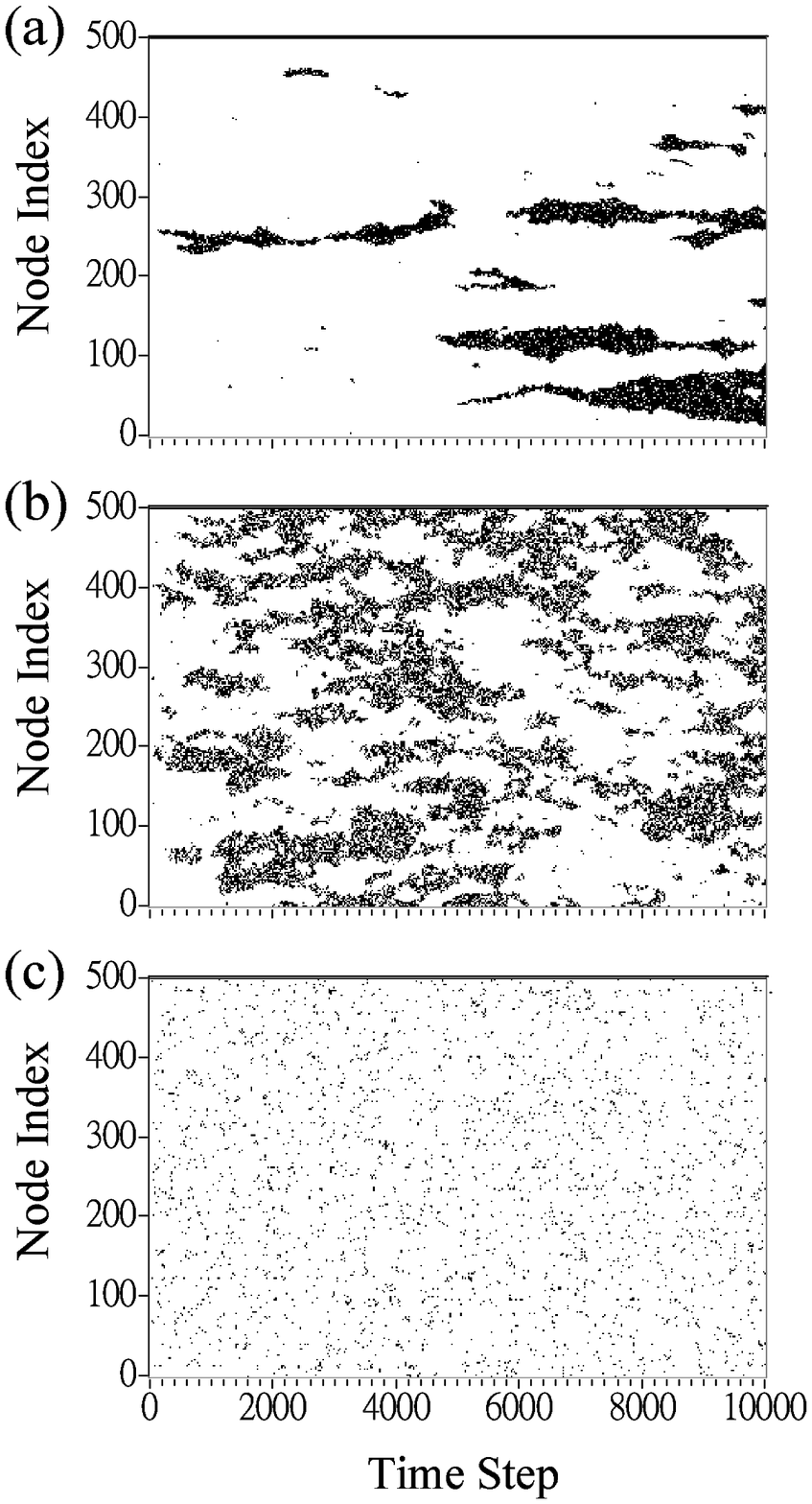}
\end{center}
\caption{Space-time distribution of agents' states for $\phi$ =
(a)0.05, (b)0.1, and (c)0.2.} \label{TE_fig1}
\end{figure}

\noindent The simplest network topology, with a range distance $k$=1 and no
short cuts, $m$=0, was first studied to elucidate the main factors in
determining the evolution caused by transmission errors.  Such a topology is
true of a 1D regular lattice, yielding nontrivial dynamics and thus
uncovering the underlying features. Some clusters consisting of agents with
the same state begin to develop during the evolution, as shown in Fig. 1(a).
Here, the transmission error, $\phi$, is set to 0.05. By increasing the
value of $\phi$ to be 0.1, as shown in Fig. 1(b), clusters converge and split
as time evolves in the plot of time-space distribution of agents' states. The
clusters developed more rapidly in the space direction, while their
persistence over time was lower than in part (a). Further increasing $\phi$
to 0.2, sizes of the clusters were much reduced, as shown in Fig. 1(c).
Indeed, increasing the transmission error amplifies the uncertainty of
information from neighbors to nodes, causing low coherence among the states
of the connected nodes. Fig. 2(a) displays the time series $S(t)$ for
$\phi=0.1$. $S(t)$ shows an irregular fluctuation after the transition from
the "all +1" state. A Fourier transform was performed on $S(t)$ to determine
the power spectrum $P(f)$, and thereby analyze the non-stationary dynamical
behavior, as shown in Fig. 2(b). In the low frequency regime, the power
spectrum has the quasi-exponential form,
\begin{equation}\label{exp_scaling}
\ln P(f)\propto -f.
\end{equation}
Surprisingly, in the high frequency regime, the tail of $P(f)$ exhibited a
novel power-law dependence on frequency,
\begin{equation}\label{power-law}
    P(f)\sim f^{-\alpha}.
\end{equation}
The dotted line represented the cumulative amplitude, $\overline{P}(f)$,
of the power spectrum, $P(f)$, defined as follows,
\begin{equation}\label{cumu_p}
    \overline{P}(f)=\int^{\infty}_{f}P(f')df'.
\end{equation}
If $P(f)$ displays a power-law scaling as in Eq.~(\ref{power-law}), then
$\overline{P}(f)$ also possesses the property of power-law scaling but with
a different exponent, $\beta=1-\alpha$. Fitting to the result showed
$\beta\simeq-1.0$ and therefore $\alpha\simeq2.0$ which , indeed, is the
prediction for the power spectrum of a random walk, $P(f)\propto1/f^{2}$
~\cite{rf12}. \bigskip

\begin{figure}
\begin{center}
\includegraphics*[scale=0.4]{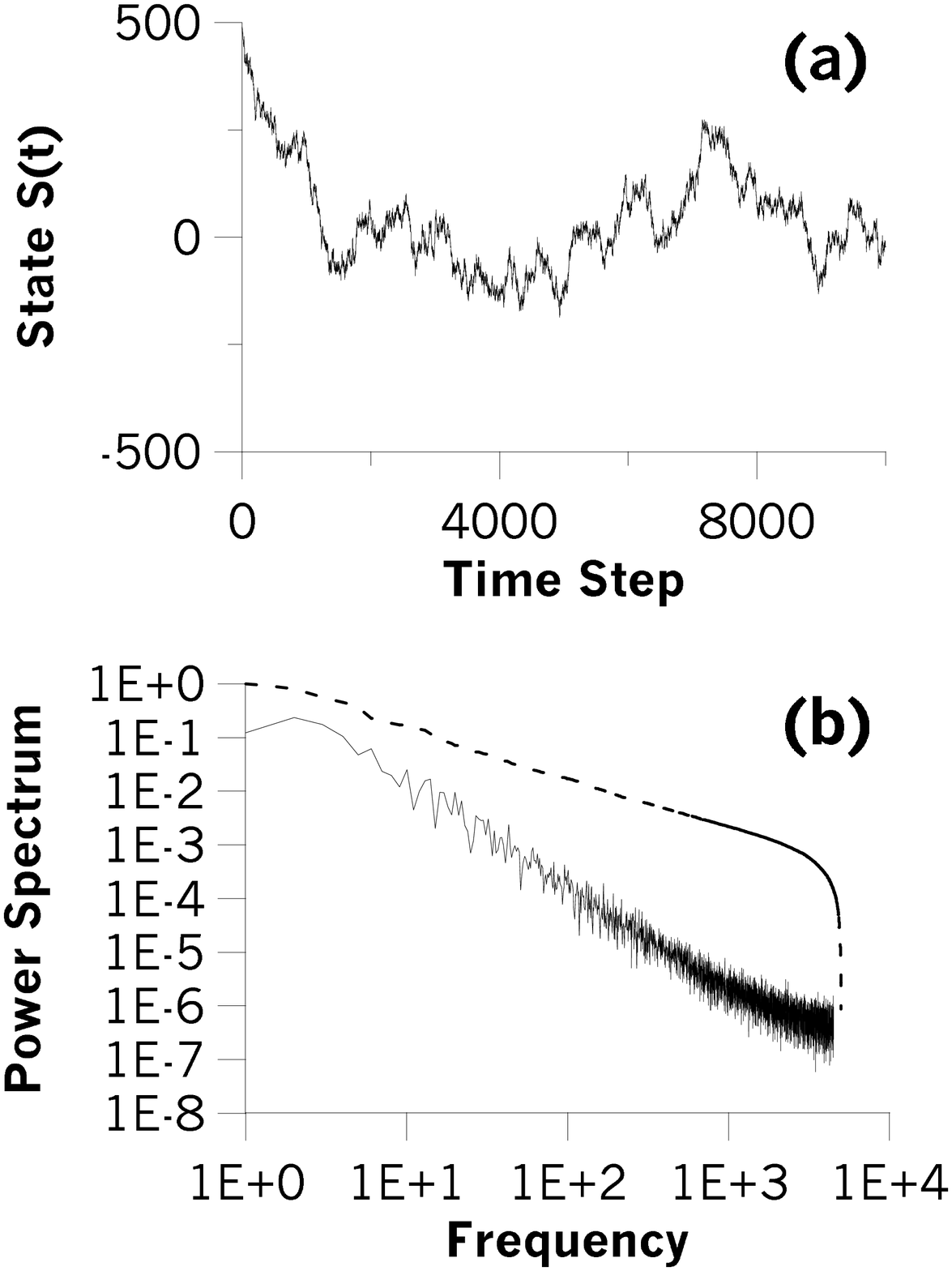}
\end{center}
\caption{(a) Time evolution of system states $S(t)$ and (b) its power
spectrum $P(f)$ (continuous line) and cumulative amplitude
$\overline{P}(f)$ (dotted line) defined in Eq.~(\ref{cumu_p}). In
the regime of power-law scaling, the exponent of
$\overline{P}(f)$,  $\beta\sim-0.9$. Thus, $\alpha\ = 1-\beta
\sim1.9$. Data used here is the same as in Fig. 1(b).} \label{TE_fig2}
\end{figure}

\noindent Simulations with different error rates were performed to further
examine the effect of the transmission error. Figure 3 shows that the
exponent of power-law scaling is almost constant over different error rates,
whereas the range of quasi-exponential decay in the low frequency regime
expands with the error rate increases, as shown in the inset of Fig. 3. The
inevitable transmission error $\phi$ yields a cutoff frequency, $f_{c}$, of
the power-law scaling in the low frequency regime. Fitting to the simulation
data shows that $f_{c}$ scales exponentially with $\phi$: $f_{c}\propto\exp
(14.38\phi)$. As stated above, the correlation between neighboring nodes is destroyed when
$\phi$ increases. Thus, the power-law scaling in the lower frequency regime
perishes first due to the disappearance of the long range correlation in
$S(t)$. For large $\phi$, power-law scaling is entirely destroyed by noise
(errors), leading to a uniform distribution of the power spectrum. However,
the simulation results presented here indicate that the exponent $\alpha$
in the power-law scaling is almost independent of the transmission error
$\phi$, implying the existence of a short time correlation in a noisy
environment and thus the possibility of predicting the behavior of the
system within short time.
\bigskip

\begin{figure}
\begin{center}
\includegraphics*[scale=0.4]{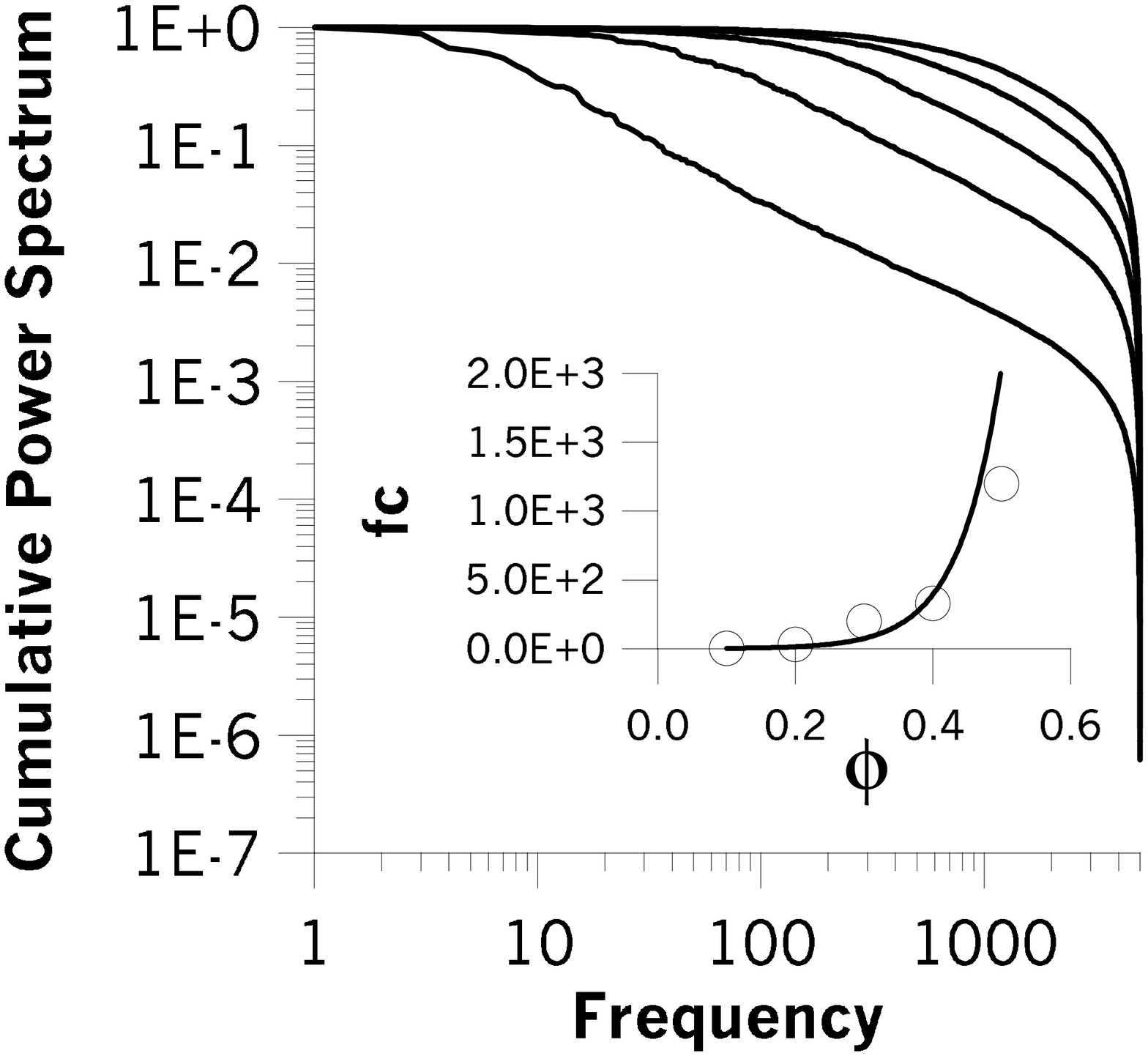}
\end{center}
\caption{Cumulative power spectral densities of $S(t)$ with different
levels of transmission error $\phi$. From up to down: $\phi$ =
0.7, 0.6, 0.5, 0.4, 0.3, 0.2, and 0.1. Inset: plot of the cut-off frequency
$f_{c}$ versus the error rate $\phi$. The fitting line (continuous) to the
data follows the form, $f_{c}\propto exp(14.38\phi)$.} \label{TE_fig3}
\end{figure}

\noindent The influence of network structures on the dynamical behavior
of our model is now further analyzed. Simulations were performed with
different levels of randomness of the small-world networks: $p$=0.01, 0.05,
0.1, 0.2, and 0.4; the other parameters were fixed: $N=500$, $k=1$, and
$\phi=0.1$. The simulation results in Fig. 4 show that the value of $\alpha$
 depends linearly on the complexity of network structures, obeying
\begin{equation}\label{a_p}
  \alpha(p)\cong 2(1-p)
\end{equation}
for $p$ less than 0.4. For large $p(>0.4)$, the networking structure is no
longer a small-world network, but like a random one. The exponent $\alpha$
saturates to zero, reflecting the disappearance of the ebb and flow of
clusters. Equation~(\ref{a_p}) was also confirmed to hold for large $N$,
$N=5000$. The size effect is weak in the model due to the fact
that the power-law scaling in the high frequency regime is substantially
contributed by the boundary interaction between clusters, causing large
fluctuation. Equations .~(\ref{exp_scaling}), ~(\ref{power-law}), and
~(\ref{a_p}) suggest the hypothesis that $P(f)$ approximately follows the
form,
\begin{equation}\label{p_form}
  P(f)\propto\frac{1}{1+[f/g(\phi)]^{\alpha}}
\end{equation} where $\ln g(\phi)\propto\phi$.
The power spectrum $P(f)$ approaches a constant for $f\ll g(\phi)$,
and exhibits power-law scaling in the high frequency regime. The two
different scaling regimes are distinguished by the cut-off
frequency, $f_{c}\sim g(\phi)$. \bigskip

\begin{figure}
\begin{center}
\includegraphics*[scale=0.4]{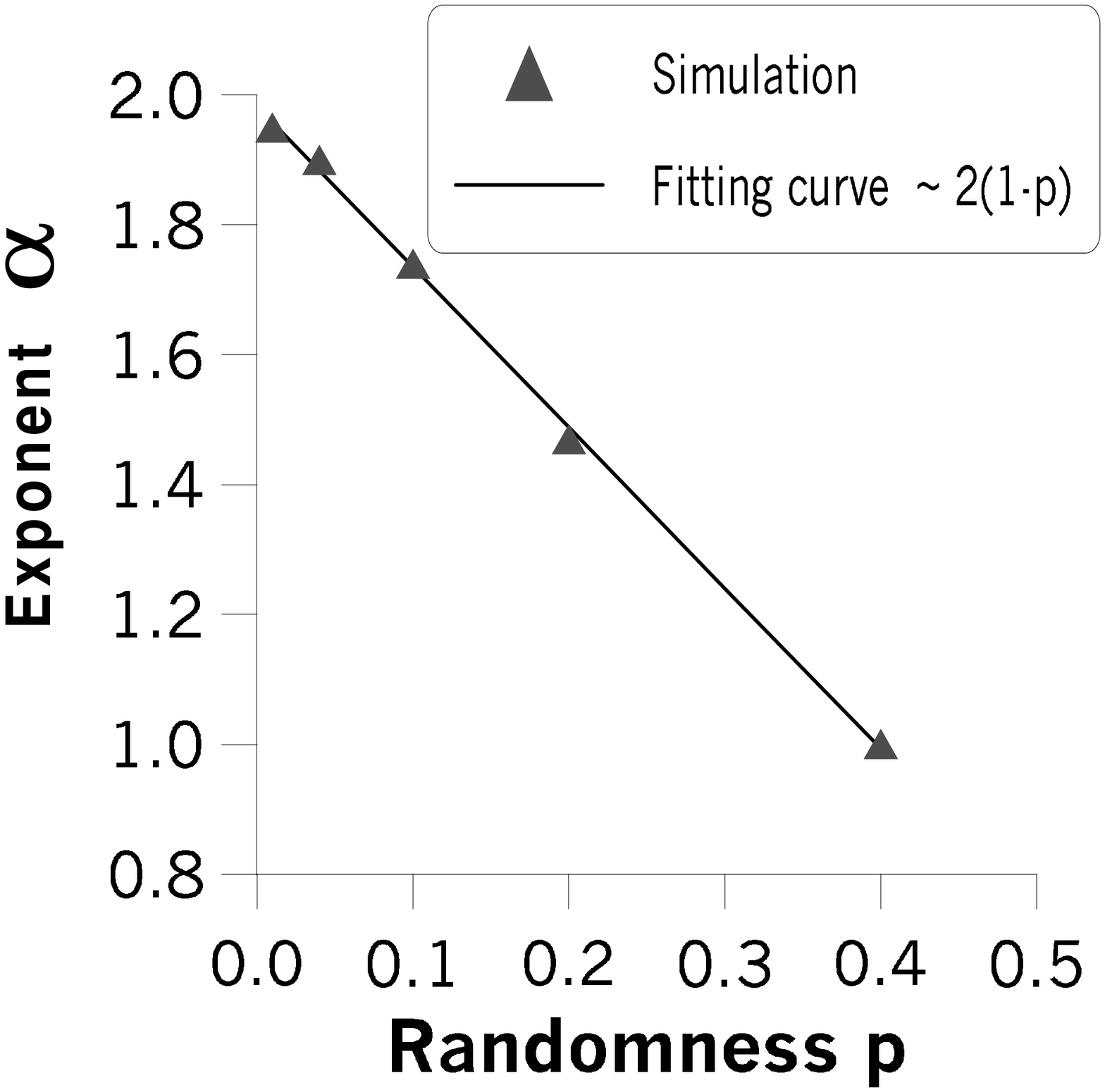}
\end{center}
\caption{Scaling exponent $\alpha$ versus randomness $p$ of
small-world network. Data points: $p$ = 0.01, 0.05, 0.1, 0.2, and 0.4}
\label{TE_fig4}
\end{figure}

\noindent Previous research into real social networks~\cite{rf1,rf4}
has revealed that the randomness of social structures range from 0.01 to
0.1, corresponding to a decrease in the value of $\alpha$ from 2.0 to 1.5
in the simulation presented here. Interestingly, recent studies
~\cite{rf11,rf12,rf13,rf14} of financial markets show that the fluctuation of
prices also exhibits power-law scaling with the exponent values ranging
from 2.0 to 1.5. The remarkable agreement of our results with empirical data
suggests that the information errors ignored in previous studies of social
phenomena may importantly participate in the appearance of power-law scaling
in complex systems. Furthermore, the power-law scaling's independence of the
system size in this model may offer a possible explanation of
the similar behavior of financial markets with different sizes but similar
social structures~\cite{rf12,rf13}. The model has been applied to examine
the scaling and fluctuating behavior of financial prices. Preliminary
results successfully resemble empirical results in which frequent large
events occur abruptly over time evolution of price turns. Detailed results
will be published elsewhere. \bigskip

\noindent In summary, in investigating the communication networks with
respect to the accuracy of transmitted information, the power spectrum of
the population difference between agents with "+1" and "-1" has been shown
to exhibit a power-law tail: $P(f)\sim f^{-\alpha}$. An important feature
is that the exponent $\alpha$ is independent of the error rate, $\phi$, but
does depend on the structure of networks. Instead, the error rate enters
only into the frequency scales and thereby determines the frequency range
over which these power laws apply. However, the exponent, $\alpha$, depends
on the intricate internal structure of networks and can serve as a structure
factor to classify the complexity of networks. Since, in pratice,
transmission error is inevitable, finite transmission error has shown to be
an important factor responsible for the common appearance of power-law
relation in complex systems, such as financial markets and opinion formation.
\bigskip

\noindent This work is partially supported by the National Science
Council, Taiwan, under the project: NSC89-2112-M-006-023.

\end{document}